\documentclass[onecolumn,showpacs]{revtex4}

\usepackage{graphicx}% Include figure files
\usepackage{dcolumn}% Align table columns on decimal point
\usepackage{bm}% bold math

\input epsf

\begin{document}

\title{\Large Emergent Universe with Exotic Matter in Brane World Scenario}

\author{\bf Ujjal Debnath$^1$\footnote{ujjaldebnath@yahoo.com ,
ujjal@iucaa.ernet.in} and Subenoy
Chakraborty$^2$\footnote{schakraborty@math.jdvu.ac.in ,
subenoyc@yahoo.co.in}}

\affiliation{$^1$Department of Mathematics, Bengal Engineering and
Science University, Shibpur, Howrah-711 103, India\\
$^2$Department of Mathematics, Jadavpur University, Kolkata 700 032, India.\\
}

\date{\today}

\begin{abstract}
In this work, we have examined the emergent scenario in brane
world model for phantom and tachyonic matter. For tachyonic matter
field we have obtained emergent scenario is possible for closed,
open and flat model of the universe with some restriction of
potential. For normal scalar field the emergent scenario is
possible only for closed model and the result is identical with
the work of Ellis et al [2], but for phantom field the emergent
scenario is possible for closed, open and flat model of the
universe with some restriction of potential.
\end{abstract}


\maketitle

In recent years to avoid the big-bang singularity, attempts have
been made in the perspective of quantum gravity (especially in
loop formalism) as well as in classical general relativity. The
idea of emergent universe is the result of the search for
singularity free inflationary model in general relativity. An
emergent universe is a model universe without a singularity, which
is ever exiting and has an almost static behaviour in the infinite
past ($t\rightarrow -\infty$) and evolves into an inflationary
stage. Also an extension of the original Lemaitre-Eddington model
is termed as emergent universe. \\

Since 1967, there was solution in GR having no big-bang
singularity. Harrison [1] described a model of a closed universe
with radiation, which as infinite past coincides with Einstein
static model. Then after a long gap (of about 40 years) similar
kind of model was discovered by Ellis et al [2, 3]. They obtained
closed model of the universe with a minimally coupled scalar field
having self-interacting potential together with a non-interacting
perfect fluid having equation of state $p=w\rho$ ($-\frac{1}{3}\le
w \le 1$). Instead of finding exact analytical solutions, they
studied the asymptotic behaviour to characterize the emergent
model. Subsequently, Mukherjee et al [4] obtained solutions of
semiclassical Einstein equations (for the Starobinsky model [4])
for flat FRW space-time and examined the features of emergent
universe. Then a general framework for emergent universe model
showing non-singular (i.e., geodesically complete) inflationary
solution was proposed by Mukherjee et al [5] where a part of the
matter is in exotic form. Also Campo et al [6] have studied an
emergent universe model for self-interacting Brans-Dicke theory.
Recently, Banerjee et al [7] have shown a model of emergent
universe in brane world scenario, while Debnath [8] has presented
an emergent universe model for exotic matter in
the form of phantom or tachyonic field.\\

In this work, we extend the idea of Debnath [8] to brane scenario.
According to Randall and Sundram [9] our universe is assumed to be
a 3-brane embedded in a 5D bulk space-time. All standard model
fields are confined on the brane, only gravity can propagate in
the whole space-time. The Einstein equations on the brane get
modified due to the existence of extra dimension. In RS II [10]
model the effective equations of motion on the 3-brane embedded in
5D bulk having $Z_{2}$-symmetry are given by

\begin{equation}
^{(4)}G_{\mu\nu}=-\Lambda_{4}q_{\mu\nu}+\kappa^{2}_{4}\tau_{\mu\nu}+\kappa^{4}_{5}\Pi_{\mu\nu}-E_{\mu\nu}
\end{equation}

where

\begin{equation}
\kappa^{2}_{4}=\frac{1}{6}~\lambda\kappa^{4}_{5}~,
\end{equation}

\begin{equation}
\Lambda_{4}=\frac{1}{2}~\kappa^{2}_{5}\left(\Lambda_{5}+\frac{1}{6}~\kappa^{2}_{5}\lambda^{2}\right)
\end{equation}
and
\begin{equation}
\Pi_{\mu\nu}=-\frac{1}{4}~\tau_{\mu\alpha}\tau^{\alpha}_{\nu}+\frac{1}{12}~\tau\tau_{\mu\nu}+\frac{1}{8}~
q_{\mu\nu}\tau_{\alpha\beta}\tau^{\alpha\beta}-\frac{1}{24}~q_{\mu\nu}\tau^{2}
\end{equation}

and $E_{\mu\nu}$ is the electric part of the 5D Weyl tensor. Here
$\kappa_{5},~\Lambda_{5},~\tau_{\mu\nu}$ and $\Lambda_{4}$ are
respectively the 5D gravitational coupling constant, 5D
cosmological constant, the brane tension (vacuum energy), brane
energy-momentum tensor and effective 4D cosmological constant.
Choosing our 3-brane as FRW universe i.e.,

\begin{equation} ds^{2} \equiv q_{\mu\nu}dx^{\mu}dx^{\nu} = dt^{2}-a^{2}(t)\left[\frac{dr^{2}}{1-kr^{2}}+
r^{2}(d\theta^{2}+\sin^{2}\theta d\phi^{2})\right]
\end{equation}

the explicit form of the above modified Einstein equations are

\begin{equation}
3H^{2}+\frac{3k}{a^{2}}=\Lambda_{4}+\kappa^{2}_{4}\rho+\frac{\kappa^{2}_{4}}{2\lambda}~\rho^{2}+\frac{6}{\lambda
\kappa^{2}_{4}}\cal{U}
\end{equation}
and
\begin{equation}
2\dot{H}+3H^{2}+\frac{k}{a^{2}}=\Lambda_{4}-\kappa^{2}_{4}p-\frac{\kappa^{2}_{4}}{2\lambda}~\rho
p-\frac{\kappa^{2}_{4}}{2\lambda}~\rho^{2}-\frac{2}{\lambda
\kappa^{2}_{4}}\cal{U}
\end{equation}

where matter in the bulk is chosen as the perfect fluid having
energy-momentum tensor

\begin{equation}
\tau_{\mu\nu}=(\rho+p)u_{\mu}u_{\nu}+pq_{\mu\nu}
\end{equation}

The matter density $\rho$ on the brane satisfies the continuity
equation

\begin{equation}
\dot{\rho}+3H(\rho+p)=0
\end{equation}

and the dark radiation $\cal{U}$ obeys

\begin{equation}
\dot{\cal U}+4H{\cal U}=0
\end{equation}

In the present work, we consider exotic matter on the brane in the
form of phantom field or tachyonic field and examine the
possibility of an emergent universe. For emergent model of the
universe, the scale factor may be chosen as [5, 8]

\begin{equation}
a = a_{0}\left(\beta+e^{\alpha t}\right)^{n}
\end{equation}

where $a_{0},~\alpha,~\beta$ and $n$ are positive constants. So
the Hubble parameter and its derivatives are given by

\begin{equation}
H=\frac{n\alpha e^{\alpha t}}{\left(\beta+e^{\alpha t}\right)}~,~
\dot{H}=\frac{n\beta\alpha^{2}e^{\alpha t}}{\left(\beta+e^{\alpha
t}\right)^{2}}~,~\ddot{H}=\frac{n\beta\alpha^{3}e^{\alpha
t}(\beta-e^{\alpha t})}{\left(\beta+e^{\alpha t}\right)^{3}}
\end{equation}

Here $H$ and $\dot{H}$ are both positive, but $\ddot{H}$ changes
sign at $t=\frac{1}{\alpha}~\text{log}\beta$. Thus $H,~\dot{H}$
and $\ddot{H}$ all tend to zero as $t\rightarrow -\infty$. On the
other hand as $t\rightarrow \infty$ the solution gives
asymptotically a de Sitter universe.\\

For the above choice of scale factor, the deceleration parameter
$q$ can be simplified to the form

\begin{equation}
q=-1-\frac{\beta}{ne^{\alpha t}}
\end{equation}

$\bullet$ {\bf Normal or Phantom field:} The phantom field has
negative kinetic term so that the ratio between pressure and
energy density is always less than $-1$. The explicit form of
energy density and pressure are [11]

\begin{equation}
\rho=\rho_{\phi}=\frac{\delta}{2}~\dot{\phi}^{2}+V(\phi)
\end{equation}
and
\begin{equation}
p=p_{\phi}=\frac{\delta}{2}~\dot{\phi}^{2}-V(\phi)
\end{equation}

where $\delta=+1$ corresponds to usual scalar field and
$\delta=-1$ represents phantom scalar field. Now subtracting field
equations (6) and (7) and using the expressions for $\rho$ and $p$
from (14) and (15), we have

\begin{equation}
\kappa^{2}_{4}\delta \dot{\phi}^{2}= -2\dot{H}+\frac{2k}{a^{2}}
-\frac{\kappa^{2}_{4}}{\lambda}\left(\frac{\dot{\phi}^{4}}{4}-V^{2}(\phi)
\right)-\frac{\kappa^{2}_{4}}{\lambda}~\rho^{2}-\frac{8}{\lambda
\kappa^{2}_{4}}\cal{U}
\end{equation}

\begin{figure}
\includegraphics[height=2in]{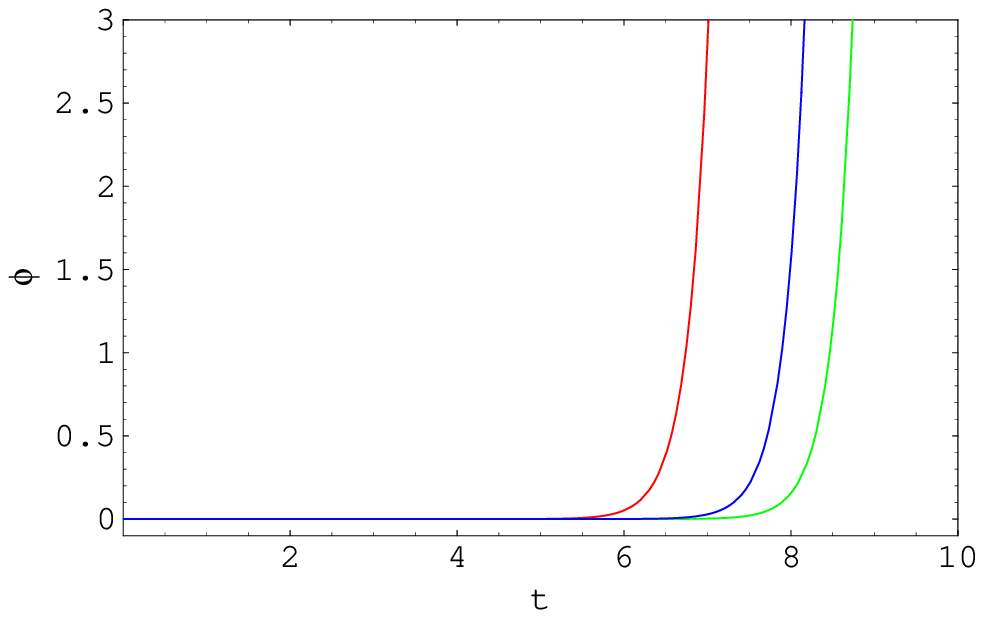}~~~~
\includegraphics[height=2in]{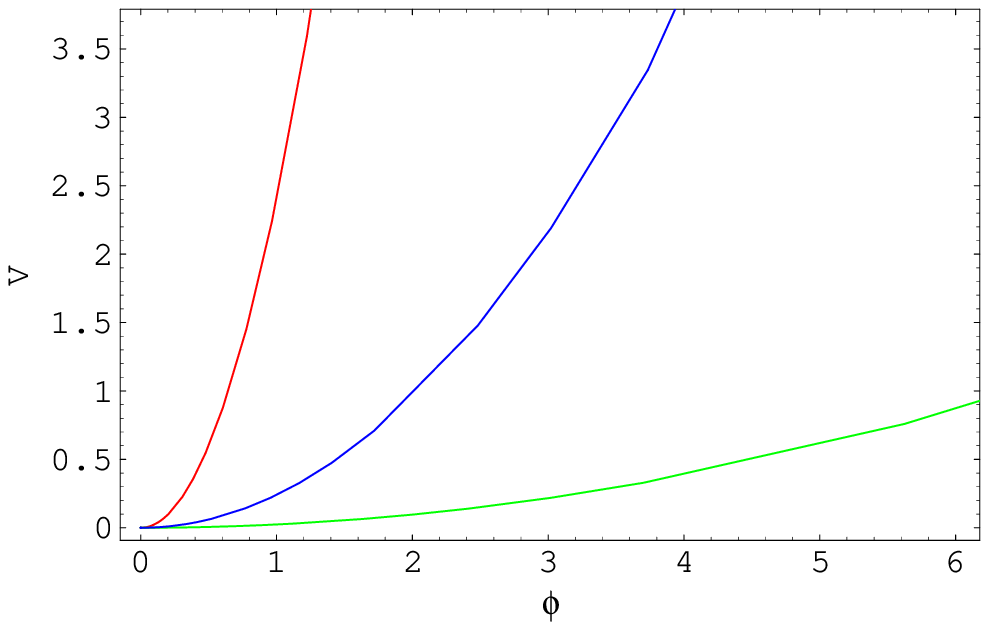}\\
\vspace{1mm} ~~~~~~~~~~~~Fig.1~~~~~~~~~~~~~~~~~~~~~~~~~~~~~~~~~~~~~~~~~~~~~~~~~~~~~~~~~~~~~~~~~~~~~~~~~~~~~~~~~~Fig.2\\

\vspace{6mm} Fig. 1 represents the variation of $\phi$ against $t$
for normalizing the constants $\alpha=2,~\beta=4$ and Fig. 2 shows
the variation of $V$ against $\phi$ for normalizing the constants
$\alpha=2,~\beta=4,~n=4,~a_{0}=1,\kappa=1,\lambda=10,\Lambda=0$ in
phantom field for closed (red line), flat (blue line) and open
(green line) universe.

\vspace{7mm}
\end{figure}

This is a quadratic in $\dot{\phi}^{2}$ and we get

\begin{equation}
\dot{\phi}^{2}=\frac{-B+\sqrt{B^{2}-4AC}}{2A}
\end{equation}

with
$A=\frac{\kappa^{2}_{4}}{2\lambda}>0,~B=\kappa^{2}_{4}\delta\left(
1+\frac{V(\phi)}{\lambda}\right)~,~C=2\dot{H}-\frac{2k}{a^{2}}+\frac{8}{\lambda
\kappa^{2}_{4}}\cal{U}$.\\

We shall now state the various possibilities in the form of
propositions:\\

{\bf Proposition I:} For normal scalar field the emergent scenario
is possible only for the closed model of the universe and we must
have

\begin{equation}
2\dot{H}+\frac{8}{\lambda \kappa^{2}_{4}}{\cal U}<\frac{2}{a^{2}}
\end{equation}

This clearly follows from the expression of $\dot{\phi}^{2}$ in
equation (17). Note that for normal scalar field $B>0$ and hence
to make the r.h.s. of (17) to be positive definite we must have
$C<0$. This is possible only when $k>0$ (i.e., closed model) and
we have the given restriction. Note also that here the dark
radiation term is against the formation of an emergent universe.\\

{\bf Proposition II:} For phantom field, the emergent scenario is
possible for any model (closed, open or flat) of the universe,
however for open or flat model the potential is restricted by the
relation

\begin{equation}
V(\phi)>\frac{2\sqrt{\lambda}}{\kappa_{4}}
\left(\dot{H}-\frac{k}{a^{2}}+\frac{4}{\lambda
\kappa^{2}_{4}}\cal{U}\right)^{\frac{1}{2}}-1
\end{equation}

In case of phantom scalar field ($\delta=-1$) $B<0$, so the r.h.s.
of (17) will be positive definite for all values of $C$ provided
$B^{2}>4AC$ for $C>0$ and it results the above restriction for
$V(\phi)$.\\

Fig. 1 and 2 show the variation of $\phi$ against $t$ and the
variation of $V$ against $\phi$ respectively for normalizing the
constants
$\alpha=2,~\beta=4,~n=4,~a_{0}=1,\kappa=1,\lambda=10,\Lambda=0$
for phantom field for closed (red line), flat (blue line) and open
(green line) universe. Fig. 1 shows the early time $\phi$ is
almost constant and then increases very sharply at finite time,
while $V$ continuously increases with $\phi$.\\

\begin{figure}
\includegraphics[height=2in]{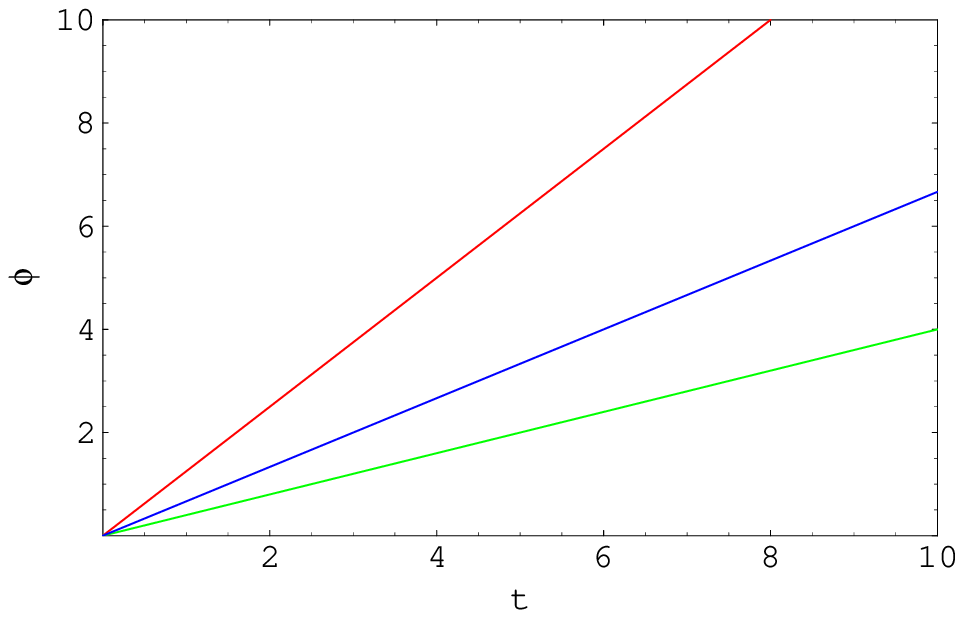}~~~~
\includegraphics[height=2in]{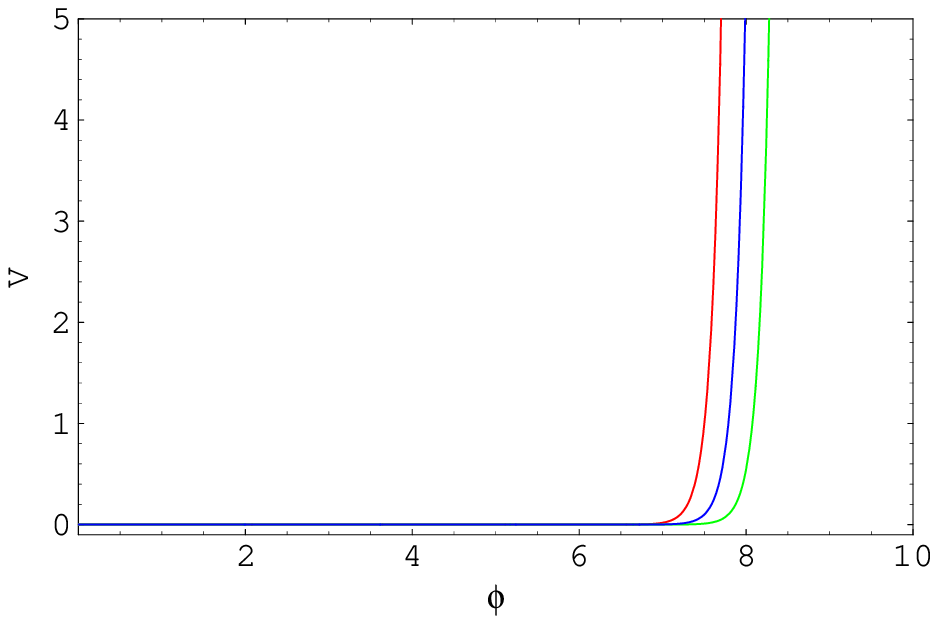}\\
\vspace{1mm} ~~~~~~~~~~~~Fig.3~~~~~~~~~~~~~~~~~~~~~~~~~~~~~~~~~~~~~~~~~~~~~~~~~~~~~~~~~~~~~~~~~~~~~~~~~~~~~~~~~~Fig.4\\

\vspace{6mm} Fig. 3 represents the variation of $\phi$ against $t$
for normalizing the constants $\alpha=2,~\beta=4$ and Fig. 4 shows
the variation of $V$ against $\phi$ for normalizing the constants
$\alpha=2,~\beta=4,~n=4,~a_{0}=1,\kappa=1,\lambda=10,\Lambda=0$ in
phantom tachyonic field for closed (red line), flat (blue line)
and open (green line) universe.

\vspace{7mm}
\end{figure}

$\bullet$ {\bf Tachyonic field:} The Lagrangian for the tachyonic
field $\psi$ having potential $U(\psi)$ is given by [12]

\begin{equation}
{\cal L}=-U(\psi)\sqrt{1-\epsilon \dot{\psi}^{2}}
\end{equation}

where as before $\epsilon=+1$ represents the normal tachyon and
$\epsilon=-1$ corresponds to phantom tachyon. The expressions for
energy density and pressure are

\begin{equation}
\rho=\rho_{\psi}=\frac{U(\psi)}{\sqrt{1-\epsilon \dot{\psi}^{2}}}
\end{equation}
and
\begin{equation}
p=p_{\psi}=-U(\psi)\sqrt{1-\epsilon \dot{\psi}^{2}}
\end{equation}

Due to the above complicated forms for $\rho$ and $p$. Similar to
the previous case, it is not possible to find an expression for
$\dot{\psi}^{2}$ from the field equations (or any combination of
them). However, from the field equation (7) we may write

\begin{equation}
1-\epsilon
\dot{\psi}^{2}=\frac{1}{\kappa^{2}_{4}\rho\left(1+\frac{\rho}{\lambda}
\right)}
\left[2\dot{H}+3H^{2}+\frac{k}{a^{2}}-\Lambda_{4}+\frac{\kappa^{2}_{4}}{2\lambda}~\rho^{2}+\frac{2}{\lambda
\kappa^{2}_{4}}\cal{U}\right]
\end{equation}

The expression for the Lagrangian or $\rho$, $p$ shows that the
r.h.s. of equation (23) must be positive definite and this is
possible for all values of $k$ i.e., for open, closed and flat
model of the universe. Thus it is possible to have emergent
scenario for any model (closed, open or flat) of the universe with
tachyonic field as the matter content, provided

\begin{equation}
2\dot{H}+3H^{2}+\frac{\kappa^{2}_{4}}{2\lambda}~\rho^{2}+\frac{2}{\lambda
\kappa^{2}_{4}}{\cal U}>\Lambda_{4}-\frac{k}{a^{2}}
\end{equation}

Fig. 3 and 4 show the variation of $\phi$ against $t$ and the
variation of $V$ against $\phi$ respectively for normalizing the
constants
$\alpha=2,~\beta=4,~n=4,~a_{0}=1,\kappa=1,\lambda=10,\Lambda=0$
for phantom tachyonic field for closed (red line), flat (blue
line) and open (green line) universe. Here $\phi$ increases with
constant slope, while $V$ remains almost constant for a wide range
of $\phi$ and then increases very sharply.\\

Thus in the present work, we have shown emergent universe in brane
scenario with 5D bulk energy in the form of a cosmological
constant and both standard and exotic matter in brane. For normal
scalar field, the emergent scenario is possible only for closed
model and the result is same as that of Ellis et al [2] and
Debnath [8] in GR. In brane world scenario, the emergent model is
possible with phantom matter for all values of the curvature
parameter $k$ and there is a realization on the potential function
in open and flat cases. In GR, Debnath [8] has obtained emergent
model with normal tachyonic field only for closed universe but
here in brane scenario, it is possible for all $k$. Here the
correction term quadratic to the energy-momentum tensor plays a
crucial role for emergent scenario. Finally, we mention that the
emergent model of Banerjee et al [7] in brane scenario, they have
chosen the matter distribution as (i) cosmological constant in the
bulk and modified Chaplygin gas in the brane and (ii) positive
cosmological constant in the bulk and perfect fluid (effectively a
radiation equation of state at high energies) in the brane and
have obtained explicit solutions in both cases.\\

{\bf Acknowledgement:}\\

The authors are thankful to IUCAA, Pune, India for warm
hospitality where part of the work was carried out.\\

{\bf References:}\\
\\
$[1]$ E. R. Harison, {\it Rev. Mod. Phys.} {\bf 39} 862 (1967).\\
$[2]$ G. F. R. Ellis and R. Maartens, {\it Class. Quantum Grav.}
{\bf 21} 223 (2004).\\
$[3]$ G. F. R. Ellis, J. Murugan and C. G. Tsagas,
{\it Class. Quantum Grav.} {\bf 21} 233 (2004).\\
$[4]$ S. Mukherjee, B. C. Paul, S. D. Maharaj and A. Beesham,
{\it gr-qc}/0505103.\\
$[5]$ S. Mukherjee, B. C. Paul, N. K. Dadhich, S. D. Maharaj and
A. Beesham, {\it Class. Quantum Grav.} {\bf 23} 6927 (2006).\\
$[6]$ S. de Campo, R. Herrera and P. Labra$\tilde{\text{n}}$a, {\it JCAP} {\bf 0711} 030 (2007).\\
$[7]$ A. Banerjee, T. Bandyopadhyay and S. Chakraborty, {\it
Gravitation and Cosmology} {\bf 13} 290 (2007); {\it Gen. Rel. Grav.} {\bf 40} 1603 (2008).\\
$[8]$ U. Debnath, {\it Class. Quantum Grav.} {\bf 25} 205019
(2008).\\
$[9]$  L. Randall and R. Sundrum, {\it Phys. Rev. Lett.} {\bf 83}
3770 (1999).\\
$[10]$ R. Maartens, {\it Living Rev. Rel.} {\bf 7} 1 (2004); L.
Randall and R. Sundrum, {\it Phys. Rev. Lett.} {\bf 83}
4690 (1999).\\
$[11]$ B. Chang, H. Liu, L. Xu and C. Zhang, {\it Chin. Phys.
Lett.} {\bf 24} 2153 (2007).  ({\it
astro-ph}/0704.3768).\\
$[12]$ J. -g. Hao and X. -z. Li, {\it Phys. Rev. D} {\bf 68}
043510 (2003); {\it Phys. Rev. D} {\bf 68} 083514 (2003); S.
Nojiri and S. D. Odintsov, {\it Phys. Lett. B} {\bf 571} 1 (2003);
B. Gumjudpai, T. Naskar, M. Sami and S. Tsujikawa,
{\it JCAP} {\bf 0506} 007 (2005).\\

\end{document}